\documentclass[conference]{IEEEtran}
\IEEEoverridecommandlockouts
\usepackage{cite}
\usepackage{amsmath,amssymb,amsfonts}
\usepackage{algorithmic}
\usepackage{graphicx}
\usepackage{textcomp}
\usepackage{xcolor}
\usepackage{xspace}
\usepackage{float}
\usepackage{amssymb, amsmath}
\usepackage[T1]{fontenc}
\usepackage{cite}
\usepackage{orcidlink}
\usepackage{academicons}
\def\BibTeX{{\rm B\kern-.05em{\sc i\kern-.025em b}\kern-.08em
    T\kern-.1667em\lower.7ex\hbox{E}\kern-.125emX}}
\begin{document}

\title{Meta-Polyp: a baseline for efficient Polyp segmentation \\
}

\author{
\IEEEauthorblockN{Quoc-Huy Trinh} \orcidlink{0000-0002-7205-3211}
\IEEEauthorblockA{\textit{Faculty of Information Technology, University of Science, VNU-HCM} \\
\textit{Vietnam National University, Ho Chi Minh City}\\
\textit{SpexAI GmbH, Dresden, Germany} \\
20120013@student.hcmus.edu.vn}
}

\maketitle

\begin{abstract}
In recent years, polyp segmentation has gained significant importance, and many methods have been developed using CNN, Vision Transformer, and Transformer techniques to achieve competitive results. However, these methods often face difficulties when dealing with out-of-distribution datasets, missing boundaries, and small polyps. In 2022, MetaFormer was introduced as a new baseline for vision, which not only improved the performance of multi-task computer vision but also addressed the limitations of the Vision Transformer and CNN family backbones. To further enhance segmentation, we propose a fusion of MetaFormer with UNet, along with the introduction of a Multi-scale Upsampling block with a level-up combination in the decoder stage to enhance the texture, also we propose the Convformer block base on the idea of the MetaFormer to enhance the crucial information of the local feature. These blocks enable the combination of global information, such as the overall shape of the polyp, with local information and boundary information, which is crucial for the decision of the medical segmentation. Our proposed approach achieved competitive performance and obtained the top result in the State of the Art on the CVC-300 dataset, Kvasir, and CVC-ColonDB dataset. Apart from Kvasir-SEG, others are out-of-distribution datasets. The implementation link can be found at: https://github.com/huyquoctrinh/MetaPolyp-CBMS2023
\end{abstract}


\section*{Acknowledgement}
This research is supported by research funding from the Faculty of Information Technology, University of Science, Vietnam National University - Ho Chi Minh City.
\section{Introduction}
Colorectal cancer is a significant health problem that poses a serious threat to human health and society. Polyps are growths that form in the colon or rectum, and they can develop into cancer over time. Early diagnosis of polyps is a crucial aspect of preventive healthcare, as it can significantly improve the prognosis and treatment outcomes of patients with colorectal cancer \cite{colectoral}. Detecting and removing polyps before they become cancerous is essential in preventing the development of the disease. Therefore, early polyp diagnosis is very crucial. It can prevent the progression of colorectal cancer and its widespread impact on society. As polyps can develop over time and some can become cancerous, early detection and removal are critical in preventing the progression of the disease. By identifying and removing polyps early, patients have a much higher chance of a successful outcome, and the overall impact of colorectal cancer can be reduced \cite{colectoral}.

In recent years, early diagnosis plays a crucial role in the treatment of polyps and the prevention of colorectal cancer. However, despite its importance, the accuracy of early diagnosis is still limited by various external factors\cite{lack-endoscopic}. Therefore, polyp segmentation has become an integral part of the diagnostic process. In recent years, several Deep Learning approaches have demonstrated their effectiveness in segmenting polyp images, with some achieving competitive results in state-of-the-art performance. These approaches include UNet \cite{unet}, PraNet \cite{pranet}, UNet++ \cite{unet++}, and ResUNet \cite{unet++}. However, these methods often face a challenge in capturing the global information of polyp objects. While CNN models excel at capturing local information, they struggle to capture the overall shape of polyp objects, which is critical for accurate segmentation. This deficiency is a significant factor in the missed segment areas that are essential outputs for segmentation tasks \cite{trinh2020hcmus}. To address these problems, many approaches of Vision Transformer \cite{metaformer} performed promising results while they can capture global information, loss of deep supervision is also a promising result for improving the boundary feature for the segmentation result. However, the deficiency of previous methods is the parameters, also the lack of local information and also global information \cite{transformer} that the model learned can lead to the oversize of polyps or the missing texture in segmentation masks and it is still a challenge for the segmentation problem \cite{pvt}.  Moreover, the texture is not captured effectively in the preceding Upsampling layer \cite{unet} due to the loss of resolution in the upsampled output.

In late 2022, a new approach called MetaFormer was proposed as a baseline for combining CNN \cite{pvt} and Transformer models \cite{transformer}. MetaFormer \cite{metaformer} allows for the capture of both local and global information by utilizing downsampling via convolution to capture local features and a Transformer encoder to capture global features in later stages. This approach has been shown to improve performance in various tasks.

In our paper, we propose a Polyp MetaFormer that combines MetaFormer and UNet, a Multi-scale Upsampling block with our Level-up UpSampling technique. Our technique enhances the quality of texture in the decoder stage of UNet, which addresses the weakness of UNet in texture missing in the UpSampling stage and improves the segmentation results of the entire architecture. Our proposed method shows competitive results on state-of-the-art datasets, benchmarking our model against the weaknesses of other approaches. 

To summarize our contribution, there are 3 main ideas:

- We propose the MetaFormer Baseline with our proposed Convformer block as 4 stages as MetaFormer for capturing the fusion of global features and local features from the encoder stage.

- The Level-up Upsampling technique is proposed to enhance the texture missing in the Decoder stage of UNet.

- Demonstrate the effectiveness of the method on out-of-distribution datasets, and also get competitive results on the state-of-the-art.

\section{Related Work}

\subsection{Early diagnosis}
Colorectal cancer, arising from polyp growth in the colon or rectum, is a significant health concern with severe implications \cite{colectoral}. Early identification of polyps is vital in improving prognosis and treatment outcomes \cite{colectoral}. Detecting and removing polyps prior to malignancy is crucial for preventing disease progression. Early diagnosis of polyps is paramount in averting the extensive consequences of colorectal cancer \cite{endoscopic}. However, current black-box methods lack explanatory transparency, posing challenges in the medical imaging field \cite{colectoral}.
\subsection{Polyps segmentation}
Endoscopic image segmentation is a well-studied research field \cite{endoscopic}. Early research relied on handcrafted descriptors and a machine learning (ML) classifier that distinguished lesions from the background based on attributes like color, shape, texture, and edges \cite{lowperformance}. In recent years, deep learning and convolutional neural networks have led to many new segmentation techniques, such as UNet \cite{unet}. The UNet \cite{unet} model is considered groundbreaking as it was the first to introduce skip connections in the encoder-decoder architecture for medical segmentation tasks. This innovative technique allows for the combination of both shallow and deep features to improve the accuracy and reliability of the segmentation process. Since its inception, numerous studies and research have been conducted to further enhance the performance of this technique in the segmentation field. As a result, many advancements have been made, and the UNet \cite{unet} model has become a crucial tool for medical imaging professionals and researchers in their pursuit of more accurate and efficient segmentation methods \cite{resunet,jha2019resunet++}. 

\subsection{MetaFormer}
It has been observed that the abstracted architecture of the Transformer, known as MetaFormer \cite{metaformer}, plays a crucial role in achieving high levels of performance. This innovative architecture has demonstrated its effectiveness in various applications, particularly in natural language processing (NLP) and image recognition. By leveraging the powerful capabilities of MetaFormer \cite{metaformer}, researchers and developers have been able to achieve competitive results and make significant advancements in their respective fields.

\section{Methods}

\subsection{General architecture}
\begin{figure*}[ht]
    \centering
    \includegraphics[width=0.8\linewidth]{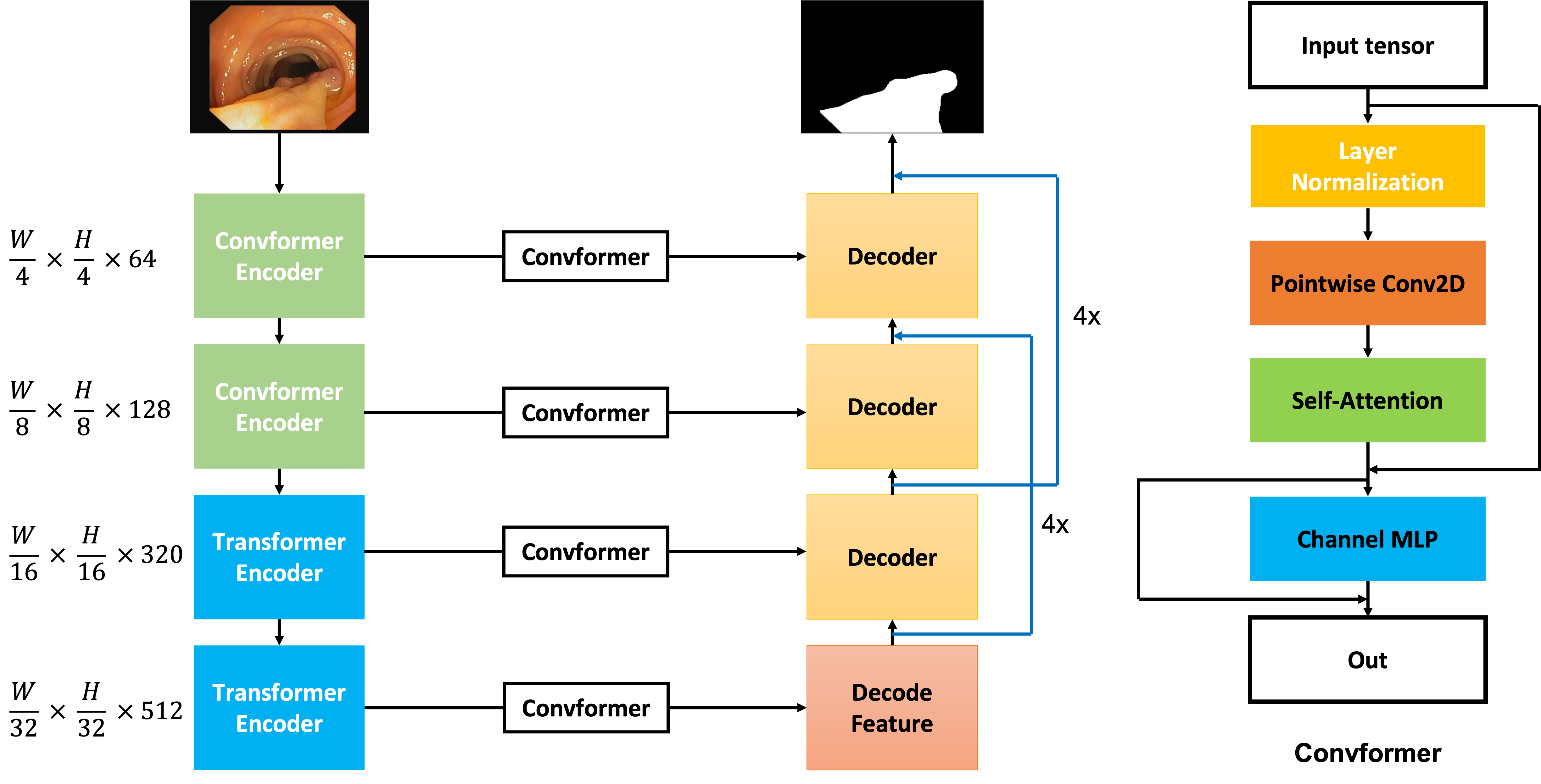}
    \caption{General architecture of Polyp MetaFormer block}
    \label{fig:archi}
\end{figure*}

We have developed a network that builds on Encoder-Decoder architecture with modifications such as the combination of the ConvFormer and Transformer blocks from the MetaFormer baseline in the encoder stage. In addition, we have proposed the use of the Level-up Upsampling stage and use the Multi-scale Upsampling block to improve the performance of the Up-Sampling layer in the decoder stage. The full architecture is described in the \ref{fig:archi}. The input $X \in R^{Width \times Height \times 3}$ of the architecture has shape $Width \times Height \times 3$, and the encoder extracts the feature $X_{i} \in R^{\frac{Width}{2^{i+1}} \times \frac{Height}{2^{i+1}} \times F_{i}}$ where $Filters_{i} \in \{64, 128, 320, 512\}; i \in \{1 , 2, 3 , 4\}$ is the filter at step $i$ of the encoder and the decoder stage. Whereas, in the decoder stage, although the feature is decoded 2 times in each step through Convolution Transpose 2D, the feature at the $i$ step is also decoded 4 times by our Multi-scale Upsampling block and then it is merged to the $i+2$ step for enhancing the feature while Upsampling the previous features. From then, the decoder stage with generate the mask with the shape $Width \times Height \times Filters$, then a 
\textit{Convolution layer $1 \times 1$} is applied to map the feature map from 64 filters to 1 filter. In the initial two stages, the emphasis lies on acquiring significant local features, which is why the Convformer Encoder is employed. Conversely, in the subsequent stages, the overall information pertaining to the object becomes more crucial. Therefore, the Transformer Encoder is incorporated in the last two stages of the model to capture the global context effectively. 

\subsection{ConvFormer Encoder}
The MetaFormer baseline \cite{metaformer} investigates how existing token mixers can achieve exceptional performance. Rather than inventing new token mixers, our work relies on the MetaFormer architecture. The ConvFormer encoder in the MetaFormer follows a four-step process. The first step involves generating token mixers, achieved through Depthwise Convolution and Separable Convolution for the creation of these mixers.

\begin{equation}
    \centering
    Convolutions(X) = Conv_{pw2}(Conv_{dw}(\sigma(Conv_{pw1}(X))))
    \label{equa1}
\end{equation}

\begin{equation}
    \centering
    X' = X' + Convolutions(Norm(X))
    \label{equa:2}
\end{equation}

\begin{equation}
    \centering
    X'' = X'' + \sigma(Norm(X')W1)W2
    \label{equa:3}
\end{equation}

The equation.\ref{equa1} which is mentioned in the \cite{metaformer}, the $Conv_{pw}$ at $i$ is the Convolution pointwise, while $Conv_{dw}$ denotes the Depth-wise Convolution. Then for the next stages, the output of the equation.\ref{equa1} is normalized before the skip connection is applied to the output, and the demonstration follows by the equation.\ref{equa:2}. The output then is learnable by $W1$ and $W2$ through the Channel MLP layer, which is demonstrated by the equation.\ref{equa:3}.In addition, in the equation.\ref{equa:3}, the $\sigma(.)$ denotes the activation function that is used in the ConvFormer block. The use of the Convformer concept in the encoder is helping the model focus on learning the important texture 

\subsection{Convformer Block}
In the Convformer block, by the idea of MetaFormer \cite{metaformer}, we create our module Convformer block (in fig.\ref{fig:archi}), which is different from the Convformer encoder, but have the idea of the previous Convformer block from MetaFormer \cite{metaformer}, that can capture the global information, also we include the local feature by the Pointwise Convolution, then the Self-Attention mechanism \cite{self-attention} is applied, in this case, the weights $W \in R^{Width \times Height \times Filter}$ is kept, this weight helps to generate the attention mask, which helps model learn the crucial information from the local information, therefore, the local information is added with the attention mask, and a Channel MLP layer is applied to help model learn the fusion of the local information and the crucial information of the local information.

\subsection{Transformer Encoder}
 The Transformer encoder shares a similar concept with the ConvFormer encoder, but with a different token mixer. Instead of using Convolution Block to create the token mixer, the Transformer block uses a classic self-attention mechanism to create an attention mask, which is used as the token mixer. The self-attention mechanism allows the model to attend to different parts of the input sequence and identify relevant features. The attention mask is generated based on the similarity between the input tokens and is used to weight the contribution of each token to the final output. This mechanism allows the model to capture long-range dependencies and contextual information.
\begin{equation}
    \centering
    X' = X' + Self Attention(Norm(X))
    \label{equa:4}
\end{equation}

\begin{equation}
    \centering
    X'' = X'' + \sigma(Norm(X')W1)W2
    \label{equa:5}
\end{equation}
In the equation.\ref{equa:4}, Self Attention is presented as the self-attention mechanism, and the output will follow the equation.\ref{equa:5}, which is the skip connection applied. Then the output is also learnable by two parameters $W1$ and $W2$ from the Channel MLP layer.

\subsection{Level-Upsampling technique}

For the Upsampling block, there are 2 stages in this block, the first stage is the Feature extraction stage, and the second stage is the Upsampling stage. 

\begin{equation}
    \centering
    X_{decoded} = Conv(UpSampling(X))
    \label{equa:7}
\end{equation}

\begin{equation}
    \centering
    X' = \sigma((X' + X_{decoded}))
     \label{equa:8}
\end{equation}

Equation.[\ref{equa:7}, \ref{equa:8}] describes the Multi-scale block with $X \in R^{W \times H \times Filters}$ as the input tensor. In this block, we utilize the Convolution module feature by extracting the input tensor by convolution layers, this one can be used with the others convolution kernel size or the others convolution components, then, the skip connection is used, the $\sigma(.)$ is the activation for the output of the step.

\section{Experimental evaluation}
\subsection{Dataset}

In the experiment, we follow the merged dataset between the ClinicDB dataset and the Kvasir-SEG dataset \cite{jha2020medico} which is mentioned in the PraNet \cite{pranet} experiment setup, and also this training set is widely used in various experiments on the later methods. This dataset contains 2 subsets: Kvasir-SEG \cite{jha2020medico} (900 train samples) and CVC-ClinicDB \cite{cvc300} (550 train samples).

For benchmarking, we choose 4 datasets: Kvasir-SEG \cite{jha2020medico}, ColonDB \cite{colondb}, CVC300 \cite{cvc300} and the Etis \cite{etis} dataset for the benchmarking. In those 4 datasets, apart from the Kvasir-SEG \cite{jha2020medico} dataset, others are out of the distribution datasets.

\begin{figure}[ht]
    \centering
    \includegraphics[height=1.9cm]{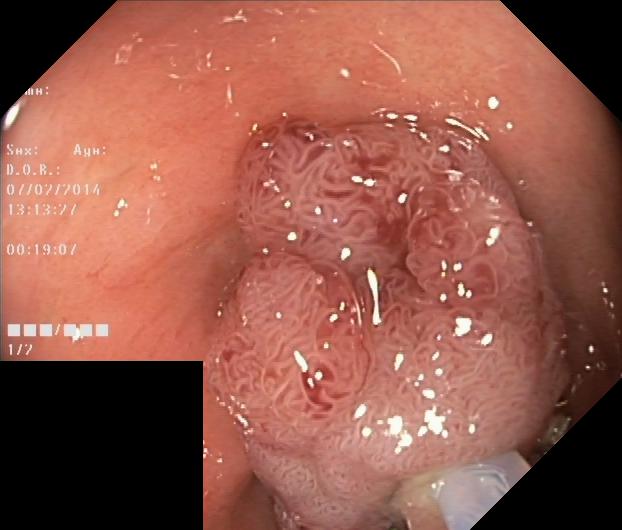}
    \includegraphics[height=1.9cm]{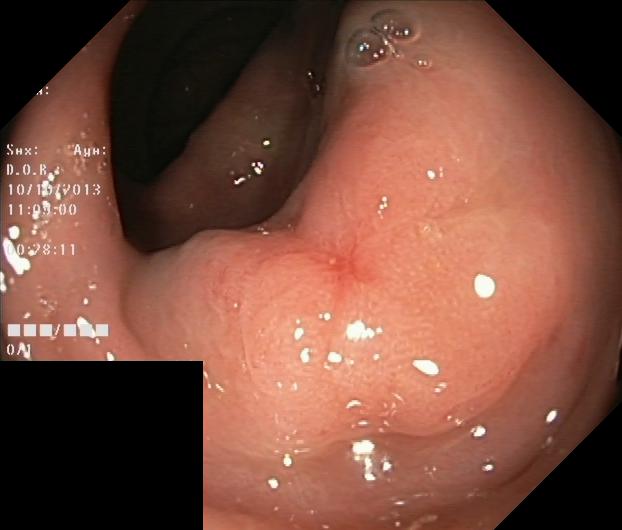}
    \includegraphics[height=1.9cm]{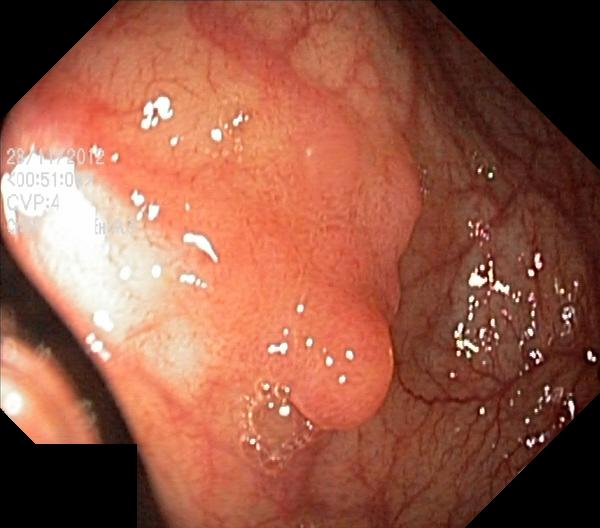}
    \includegraphics[height=1.9cm]{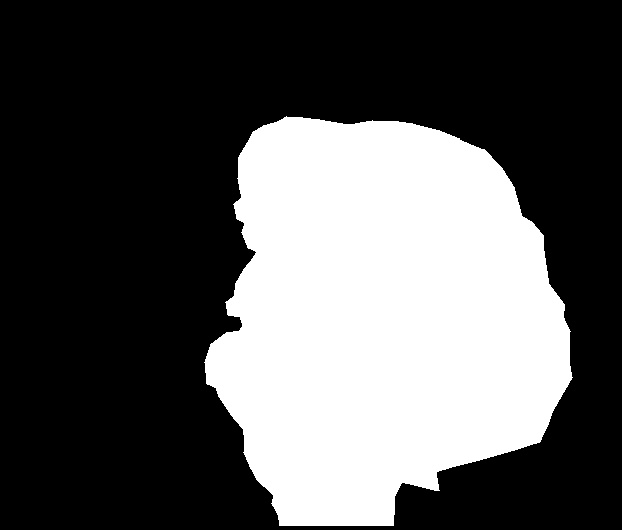}
    \includegraphics[height=1.9cm]{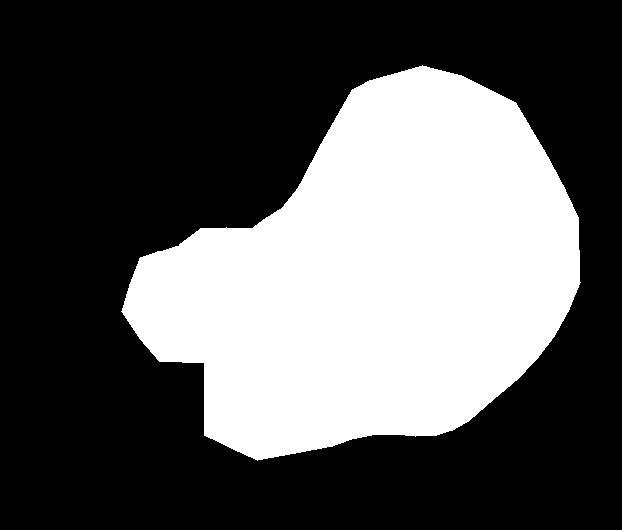}
    \includegraphics[height=1.9cm]{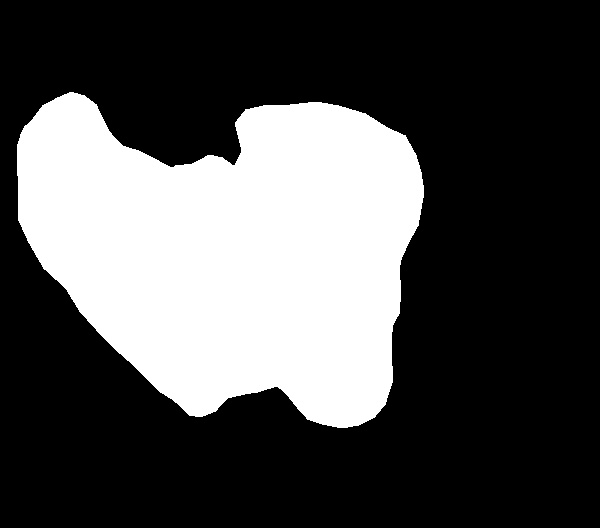}
    \caption{Polyps and corresponding masks from Kvasir-SEG}  
    \label{fig:images}
\end{figure}

For research and study purposes, we split the merged dataset into 3 parts, one for training, one for validating, and one for testing; and do experiments on all these 3 parts. The training, validating, and testing make up 60\%, 20\%, and 20\%, respectively, this data  split method is used for evaluating our model before benchmarking on various datasets. 
\subsection{Augmentation}
While training the model, we also use the Augmentation technique to improve the number of data in the dataset. Augmentation also creates a beneficial impact on the domain of the dataset by making the distribution of the data more complex \cite{augmentation}.
To enrich the dataset, we propose some augmentation methods. We use Center Crop \cite{advanceaug}, Random Rotate \cite{advanceaug}, GridDistortion \cite{advanceaug}, Horizontal \cite{advanceaug}, and Vertical Flip to improve the quantity of the dataset. Moreover, some advanced augmentation methods were applied to improve the distribution of the feature in the data sample.
\begin{figure} [H]
    \centering
    \includegraphics[height=2.1cm]{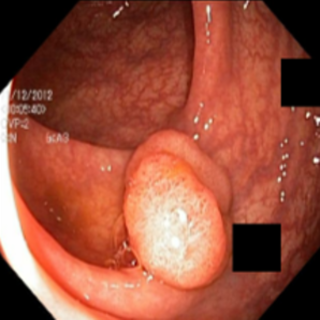}
    \includegraphics[height=2.1cm]{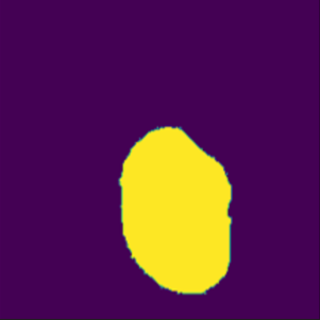}
    \caption{Example of CutOut augmentation for Kvasir-SEG}
    \label{fig:cutout}
\end{figure}
CutOut augmentation \cite{devries2017improved} is also applied in our experiments, this advanced method adds the noise, which is the area with all pixel values are 0, to the image randomly and then the mask has also added that area in the same position. Moreover, the CutMix augmentation is used, in this case, the original image is added with a patch from another image, and their corresponding masks are also combined in the same way.



\subsection{Loss function}
We utilize the Jaccard Loss Function \cite{jaacaard} with the following formula 
\begin{equation}
\centering
    Jaccard Loss(y,\hat y) = \alpha\times(1-\frac{\alpha+\sum_{c}^Cy_{c}\times\hat y_{c}}{\alpha +\sum_{c}^Cy_{c} + \hat y^{c} - y_{c}\times\hat y_{c}})
\centering 
\label{jac}
\vspace{2mm}
\end{equation}
This loss function enables the segmentation process better and can control the model's performance on the pitch of the tissues. The Jaccard Loss \cite{jaacaard} is also known as the IOU metric, with y as the true label and the predicted label being $\hat y$, these two labels are demonstrated in the one-hot vector to present classes $C$ being their length. However, to prevent the exploding gradient, there is a smoothing factor called alpha $\alpha$, which helps stabilize the training result.

\subsection{Implementation details}

All architectures were implemented using the Keras framework with TensorFlow as the backend. The input volumes are normalized to [-1, 1]. All networks were trained with described augmentation methods. We used Adam optimization \cite{kingma2014adam} with an initial learning rate of 1e-4. After that, we use the Cosine Annealing learning rate schedule to stable the training process. The smoothing factor alpha $\alpha$ in the Jaccard Loss is $0.7$. We performed
our experiment on a single NVIDIA Tesla A100 40GB. The batch size is 128, and it takes around 6 hours to train the entire dataset. Finally, we trained all the models for 300 epochs.

\subsection{Metrics}

We use IOU and Dice-Coefficient metrics to evaluate our method's performance. The metrics evaluate the ground truth mask with the predicted mask from the test dataset.

The following is the formula of mIOU \cite{iou}:
\begin{equation}
    \centering
    IOU = \frac{Area  of  Overlap}{Area of Union}
\label{iou}
\end{equation}

The Area of overlap is the common area of two predicted masks, and the Area of Union is all of the areas of two masks.

The Dice Coefficient \cite{dicecoef} which calculates the division between the common area of two masks and the union area of two masks has the following formula:
\begin{equation}
    \centering
    Dice Coefficient = \frac{2 * |X \cap Y|}{|X \cup Y|}
    \label{dicecoef}
\end{equation}

\section{Result}

\subsection{Qualitative Result}

For the relevance of our results, we also benchmark our model with previous methods from UNet \cite{unet} in 2015 to the newest method which is on the top of the state-of-the-art in early 2023 is FCB-SwinV2 Transformer \cite{fcb}. Below is the result on the Kvasir-SEG dataset\cite{jha2020medico} and CVC-300 dataset \cite{cvc300} with results that are competitive on state-of-the-art as shown in Table.\ref{tab:tab1}, and Table.\ref{tab:tab2}. In addition, to evaluate the deficiency of our approach, we propose to do experiments on the Etis dataset \cite{etis}, and the Colon-DB dataset \cite{colondb}. From the experiment, results are observed in Table.\ref{tab:tab3} and Table.\ref{tab:tab4}.
\begin{figure*}[ht]
    \centering
    \includegraphics[width=0.95\linewidth]{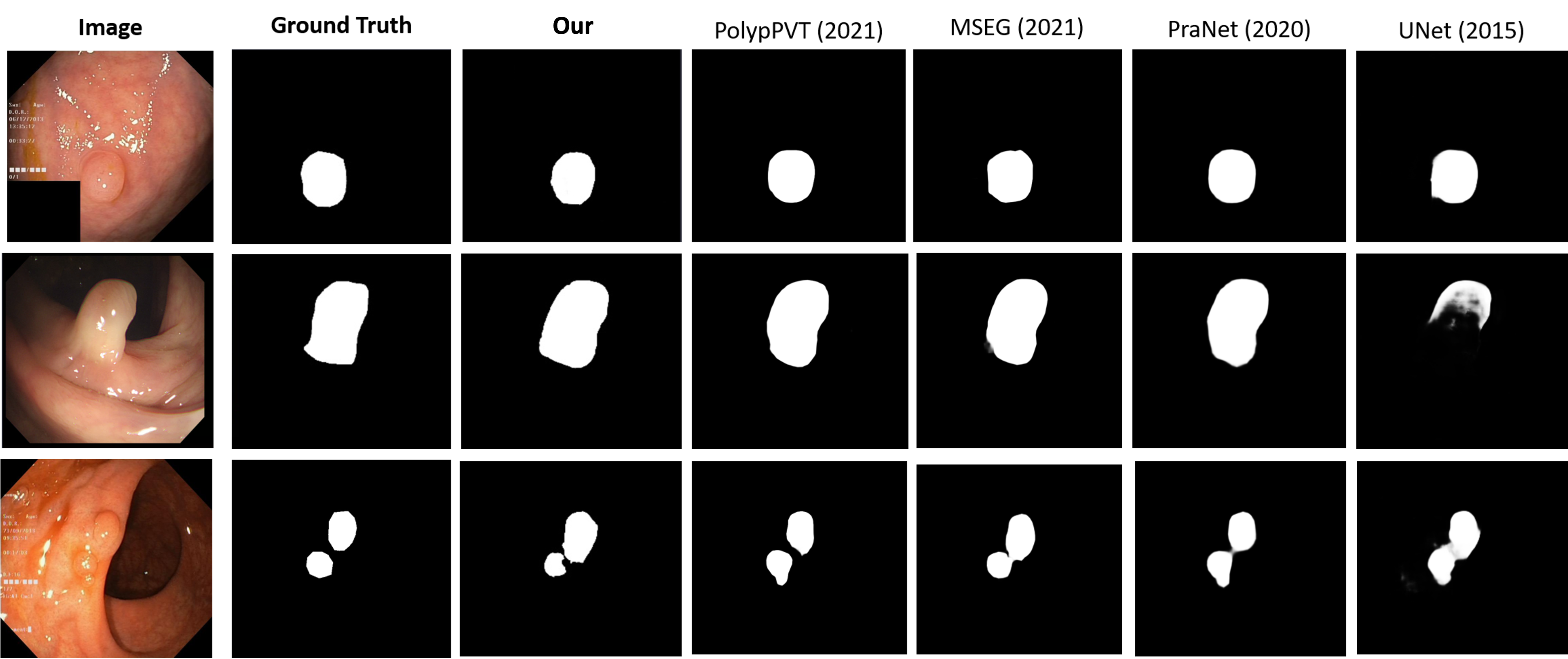}
    \caption{Comparison of results from various methods}
    \label{fig:vis}
\end{figure*}

\begin{table}[H]
\begin{center}
\begin{tabular}{|c| c | c| c|} 
 \hline
 Model & $mIoU \uparrow$ & $mDice \uparrow$ & $MAE \downarrow$ \\ [0.5ex] 
 \hline
 UNet (MICCAI 2015) \cite{unet} & 0.756 & 0.821 & 0.055 \\
 PraNet (MICCAI 2020) \cite{pranet}  & 0.840 & 0.898 & 0.030 \\
 MSEG Net (arxiv'21) \cite{mseg} & 0.857 & 0.912 & 0.025 \\

SANet (MICCAI 2021) \cite{sanet} & 0.847 & 0.904 & 0.028 \\

MSNet (MICCAI 2021) \cite{MSNet} & 0.862 & 0.907 & 0.028 \\

Polyp-PVT (arXiv'21) \cite{pvt} &  0.864 & 0.917 & 0.023\\

PEFV2 (MMM 2023) \cite{pefmmm} & 0.8163 & 0.8818 & nan \\

Polyp2Seg (MIUA 2022) \cite{polyp2seg} & 0.882 & 0.929 & 0.018\\

FCB-SwinV2 (arXiv'23) \cite{fcb} & 0.8973 & 0.942 & nan \\
\hline

\textbf{Our} & \textbf{0.921} & \textbf{0.959} & \textbf{0.004} \\

 \hline
\end{tabular}
\end{center}
\caption{Results on Kvasir Polyps dataset \cite{jha2020medico}}
\label{tab:tab1}
\end{table}

\begin{table}[H]
\begin{center}
\begin{tabular}{|c| c | c| c|} 
 \hline
 Model & $mIoU \uparrow$ & $mDice \uparrow$ & $MAE \downarrow$ \\ [0.5ex] 
 \hline
 UNet (MICCAI 2015) \cite{unet} & 0.449 & 0.519 & 0.022 \\

 PraNet (MICCAI 2020) \cite{pranet}  & 0.797 & 0.871 & 0.010 \\

SANet (MICCAI 2021) \cite{sanet} &0.815 & 0.888 & 0.008 \\

MSNet (MICCAI 2021) \cite{MSNet} & 0.807 & 0.869 & 0.010 \\

Polyp-PVT (arXiv'21) \cite{pvt} &  0.833 & 0.8900 & 0.007\\


Polyp2Seg (MIUA 2022) \cite{polyp2seg} & 0.818 & 0.890 & 0.007\\

\hline

\textbf{Our} & \textbf{0.862} & \textbf{0.926} & \textbf{0.006} \\

 \hline
\end{tabular}
\end{center}
\caption{Results on CVC300 Polyps dataset \cite{cvc300}}
\label{tab:tab2}
\end{table}

\begin{table}[H]
\begin{center}
\begin{tabular}{|c| c | c| c|} 
 \hline
 Model & $mIoU \uparrow$ & $mDice \uparrow$ & $MAE \downarrow$ \\ [0.5ex] 
 \hline
 UNet (MICCAI 2015) \cite{unet} & 0.343 & 0.406 & 0.036 \\

 PraNet (MICCAI 2020) \cite{pranet}  & 0.567 & 0.628 & 0.031 \\

SANet (MICCAI 2021) \cite{sanet} & 0.654 & 0.750 & 0.015 \\

MSNet (MICCAI 2021) \cite{MSNet} & 0.664 & 0.719 & 0.020 \\

Polyp-PVT (arXiv'21) \cite{pvt} &  0.706 & 0.787 & \textbf{0.013}\\

Polyp2Seg (MIUA 2022) \cite{polyp2seg} & \textbf{0.738} & \textbf{0.820} & 0.015\\
\hline

\textbf{Our} & \underline{0.704} & \underline{0.780} & \underline{0.035} \\

 \hline
\end{tabular}
\end{center}
\caption{Results on Etis dataset \cite{etis}}
\label{tab:tab3}
\end{table}

\begin{table}[H]
\begin{center}
\begin{tabular}{|c| c | c| c|} 
 \hline
 Model & $mIoU \uparrow$ & $mDice \uparrow$ & $MAE \downarrow$ \\ [0.5ex] 
 \hline
 UNet (MICCAI 2015) \cite{unet} & 0.449 & 0.519 & 0.061 \\

 PraNet (MICCAI 2020) \cite{pranet}  & 0.640 & 0.712 & 0.043 \\

SANet (MICCAI 2021) \cite{sanet} & 0.670 & 0.753 & 0.043 \\

MSNet (MICCAI 2021) \cite{MSNet} & 0.678 & 0.755 & 0.041 \\

Polyp-PVT (arXiv'21) \cite{pvt} &  0.727 & 0.808 & 0.031\\

Polyp2Seg (MIUA 2022) \cite{polyp2seg} & 0.727 & 0.808 & 0.031\\
\hline

\textbf{Our} & \textbf{0.790} & \textbf{0.867} & \textbf{0.009} \\

 \hline
\end{tabular}
\end{center}
\caption{Results on Colon-DB dataset \cite{colondb}}
\label{tab:tab4}
\end{table}
Overall, the Table.\ref{tab:tab2}, Table.\ref{tab:tab1}, and Table.\ref{tab:tab4}, our method achieves the state-of-the art on the Kvasir-SEG \cite{jha2020medico}, CVC-ColonDB dataset \cite{colondb}, and CVC-300 \cite{cvc300} dataset. These results evaluate the effectiveness of our proposed Convformer and Multiscale Upsampling block. However, the weakness of the method is on the small object which is illustrated by the result from the Table.\ref{tab:tab3} from the Etis dataset \cite{etis}.
\subsection{Qualitative visualization}

The fig.\ref{fig:vis} visualizes masks that are generated from our method when compared with other methods each year. The dataset that is used for this visualization is the Kvasir dataset \cite{kvasir-seg}. From the results, our method is demonstrated to improve the weakness of the previous method to identify the shape of some difficult polyp objects, however, with many polyps in the image, our method also shows a worse result than some methods in the previous year.

\subsection{Ablation study}
The experiments to compare the MetaFormer UNet with the combination of Meta-UNet and our proposed block are done to evaluate the effectiveness of our method. It seems that the Multi-scale Upsampling block has led to significant improvements in the performance of the MetaFormer UNet on the Kvasir dataset, while the Convformer block can help the model learn the local feature and also can learn the captured global feature. The increase in mean Intersection over Union (mIOU) score from 0.877 to 0.921 with the addition of the Multi-scale Upsampling block and Convformer block indicates that this technique has effectively improved the model's ability to capture fine details and produce more accurate segmentation results.  Furthermore, the best-proposed model achieving a 0.921 mIOU score on the Kvasir dataset suggests that the MetaFormer UNet with Multi-scale Upsampling block and Convformer block is capable of achieving state-of-the-art performance in this task. It would be important to consider the computational cost and other practical considerations of the Multi-scale Upsampling block, as well as potential trade-offs in other performance metrics when deciding whether to incorporate it into other models or applications.

\section{Conclusion}

In conclusion, we propose the MetaFormer baseline with UNet and our Multi-scale Upsampling block with the Level-up augmentation technique for the segmentation. Our approach is evaluated to solve the problem from the previous methods which is the lack of local features with global features that the model learned, which can help the model capture the full shape and the texture inside the mask. Moreover, our results achieve the state-of-the-art of Kvasir-SEG dataset, the CVC-ColonDB dataset, and the CVC300 dataset. This result demonstrates that our method enhances the weakness of previous methods with our proposed modules. On the other hand, there are some limitations of the MetaFormer that need to be improved such as small polyps in the segmentation or multiple polyps also make our method obtains lower performance than usual. However, this is a promising method for the medical segmentation task and can be improved in the future.

\bibliographystyle{IEEEtran}
\bibliography{myref.bib}

\end{document}